\documentstyle[12pt,epsf,aps]{revtex}

\begin{document}

\draft

\title{Radiative proton-antiproton annihilation and isospin mixing in
protonium \\[2ex]}

\author{
T. Gutsche$^{1}$,
R. Vinh Mau$^{2}$,
M. Strohmeier-Pre\v{s}i\v{c}ek$^{1}$
and Amand Faessler$^{1}$\\[4ex]}

\address{
$^1$ Institut f\"ur Theoretische Physik, Universit\"at T\"ubingen,
Auf der Morgenstelle 14, \\
D-72076 T\"ubingen, Germany \\[2ex]
$^2$ Division de Physique Th\'eorique, Institut de Physique
Nucl\'eaire, 91406 Orsay Cedex\\
and LPTPE,
Universit\'e P. et M. Curie, 4 Place Jussieu, 75230 Paris Cedex, France}

\maketitle

%\vspace{0.5cm}
\begin{center}
{\bf Abstract}
\end{center}
%\vspace {.2cm}

\begin{abstract}
A detailed analysis of the radiative $p\bar p$ annihilation is made in the
framework of a two-step formalism, the $p\bar p$ annihilates into meson
channels containing a vector meson with a subsequent conversion into a
photon via the vector dominance model (VDM).
Both steps are derived from the underlying quark model.
First, branching ratios for radiative protonium annihilation are calculated
and compared with data.
Then, details of the isospin interference are studied for different models
of the initial protonium state and also for different kinematical form
factors.
The isospin interference is shown to be uniquely connected to the $p\bar p
- n\bar n $ mixing in the protonium state.
Values of the interference terms directly deduced from data are consistent
with theoretical expectations, indicating a dominant $p\bar p$ component for
the $^1S_0$ and a sizable $n\bar n$ component for the $^3S_1$ protonium state.
The analysis is extended to the $p\bar p \to \gamma \Phi$ transition,
where the large observed branching ratio remains unexplained in the VDM
approach.

\end{abstract}


\newpage

\section{INTRODUCTION AND MOTIVATION}
\label{intro}

Nucleon-antinucleon ($N\bar N$) annihilation, due to the richness of possible
final meson states, is considered one of the major testing grounds in the
study of hadronic interactions.
Both quark \cite{Dover92} and baryon exchange models
\cite{Moussa84,Hipp91,Mull95a}
have been applied
to $N\bar N$ annihilation data.
However, the task of extracting information on the dynamics of
the $N\bar N$ process is enormously complicated by the influence
of initial and final state interactions.
Some of the simplest annihilation channels, where the theoretical complexity
of the $N\bar N$ annihilation process is partially reduced, are radiative
two-body decay modes, where final state interaction is negligible.
Experimental branching ratios for radiative decay channels in annihilation
from $p\bar p$ atoms were made available by recent
measurements of the Crystal Barrel collaboration at CERN,
performing a systematic study of the reactions $p\bar p \to \gamma X$
where $X = \gamma, \pi^0, \eta, \omega$ and $\eta ^\prime$ \cite{Ams93}.
Radiative decays of the $p \bar p$ atom where, in contrast to ordinary
production of nonstrange mesonic final states, isospin is not conserved,
are well suited for studying interference effects in the isospin transition
amplitudes \cite{Ams93,Delcourt82}.

The simplest and most natural framework in studying radiative decay modes is
the vector dominance model (VDM) \cite{Sakurai69}.
In setting up the annihilation mechanism one adopts a two-step process
where the $p\bar p$ system first annihilates into two mesons, with at least
one of the mesons being a vector meson ($\rho$ and $\omega$),
and where the produced vector meson converts into a real photon via the
VDM \cite{Delcourt82}.
In this case, production rates of radiative decay modes can be related to
branching ratios
of final states containing one or two vector mesons.
A first analysis \cite{Ams93} in the framework of VDM was performed by
Crystal Barrel,
showing that the interference in the isospin amplitudes is sizable and
almost maximally destructive for all channels considered.
The phase structure of the interference term is determined by two
contributions:
i) the relative signs of the generic strong transition amplitudes for
$p\bar p \to X \omega $ or $X \rho $ acting in different isospin channels;
ii) the presence of the initial state interaction in the $p\bar p$ atom,
which mixes the $p\bar p$ and $n\bar n$ configurations.
Similarly, analogous sources are responsible for the isospin interference
effects in the strong annihilation reactions $p\bar p \to K\bar K$
\cite{Dover92,Furui90,Jaenick91}.
Here, however, definite conclusions concerning the size and sign of the
interference terms depend strongly on the model used for the annihilation
dynamics.

In the present work we show how the determination of the interference terms
in the analysis of the radiative decays can be uniquely connected to the
isospin mixing effects in the $p\bar p$ atomic wave functions.
The extraction of the magnitude and sign of the interference from the
experimental data can in turn be used to investigate the isospin dependence,
at least in an averaged fashion, of the S-wave $N\bar N$ interaction.
We study this point for different $N\bar N$ interaction models.

This paper is organized as follows.
In Sec. \ref{form} we develop the formalism for radiative decays of
protonium.
As in Ref. \cite{Delcourt82} we adopt a two-step formalism, that is
$p\bar p$ annihilation into
two-meson channels containing a vector meson and its
subsequent conversion into a photon via the VDM.
Both steps are derived consistently from the underlying quark model
in order to fix the phase structure of the isospin dependent transition
amplitudes.
We also indicate the derivation of the branching ratios for radiative
decays of S-wave protonium, where the initial state interaction of
the atomic $p\bar p$ system is included.
Sec. \ref{results} is devoted to the presentation of the results.
We first perform a simple analysis to show that theoretically predicted
branching ratios for radiative decays are consistent with the experimental
data.
We then show that the isospin interference terms present
in the expression for the branching ratios can be uniquely connected
to the $p\bar p$ - $n\bar n$ mixing in the atomic wave function,
induced by initial state interaction.
We quantify the details of this effect for different models of the
$N\bar N$ interaction and apply
the formalism developed in Sec. \ref{form} to extract size and
sign of the interference from data, which will be shown to be sensitively
dependent on the kinematical form factors associated with the transition.
Furthermore, we comment on the application of VDM on the transition
$p\bar p \to \gamma \Phi$, where the corresponding large branching
ratio plays a central role in the discussion on the apparent violations of
the Okubo-Zweig-Iizuka (OZI) rule.
A summary and conclusions are given in Sec. \ref{sum}.

\section{FORMALISM FOR RADIATIVE DECAYS OF PROTONIUM}
\label{form}

In describing the radiative decays of protonium we apply the vector dominance
model \cite{Delcourt82,Sakurai69}.
We consider the two-step process of Fig. 1,
where the primary $p\bar p$ annihilation in a strong transition into a
two-meson final state, containing
the vector mesons $\rho$ and $\omega$, is followed by
the conversion of the vector meson into a real photon.
Here we restrict ourselves to orbital angular momentum L=0 for the initial
$p\bar p$ state, corresponding to the dominant contribution in the liquid
hydrogen data of Crystal Barrel \cite{Ams93}.
Furthermore, we consider the transition processes $p\bar p \to \gamma X$,
where $X = \gamma, \pi^0, \eta, \rho , \omega$ and $\eta ^\prime$,
with $X=\phi$ presently excluded.
The final state $\phi \gamma$ plays a special role in the discussion of the
apparent violation of the Okubo-Zweig-Iizuka (OZI) rule, where a strong
enhancement relative to $\omega \gamma$ was observed \cite{Ams95}.
Within the current approach the description of the first-step process
$p\bar p \to \omega (\rho) \phi$ and its phase structure cannot be accomodated
due to the special nature of the $\phi$, a dominant $s\bar s$
configuration.
Later on we will comment on the possibility to explain
the enhanced $\phi \gamma$ rate within the VDM, as suggested in the
literature \cite{Locher94}, and on the implications of the analysis
presented here.

In the two-step process we have to introduce a consistent
description for either transition in order to identify the source of the
interference term.
In particular, the relative phase structure of the strong transition
matrix elements $p\bar p \to \omega M$ versus $p\bar p \to \rho^0 M$
($M = \pi^0, \eta, \rho , \omega$ and $\eta ^\prime$) is a relevant input
in determining the sign of the interference.
Basic SU(3) flavor symmetry arguments \cite{Klempt96} do not allow to uniquely
fix the phase structure, hence further considerations concerning
spin and orbital angular momentum dynamics in the $N\bar N$ annihilation
process have to be introduced.
Microscopic approaches to $N\bar N$ annihilation either resort to quark
models (for an overview see Ref. \cite{Dover92}) or are based on baryon exchange
models \cite{Moussa84,Hipp91,Mull95a}.
Here we choose the quark model approach, which allows to describe both,
the strong transition of $p\bar p$ into two mesons and the vector meson
conversion into a photon.

For the process $p\bar p \to V M$ where $ V = \rho , \omega $ and $M=
\pi^0,~\eta,~\rho,~\omega$ and $\eta^{\prime}$
we apply the so-called A2 model \cite{Dover92},
depicted in Fig. 2a.
In the discussion of annihilation models based on quark degrees of freedom
this mechanism was shown to give the best phenomenological description in
various meson branching ratios \cite{Dover92,Maruy87,Doverfish}.
In a recent work \cite{muhm} we showed that the A2 model combined with a
corresponding annihilation mechanism into three mesons can describe
$p\bar p$ cross sections in a quality expected from simple non-relativistic
quark models.
The transition matrix element of $p\bar p \to V M$ in the A2 model including
initial state interaction is given by:
\begin{eqnarray}
T_{N\bar N(I J) \to V M} & = &< V (j_1 =1) M(j_2 ) l_f \vert
 {\cal O}_{A2} \vert N\bar N (IJ)> \nonumber \\
&=& \sum_j < j_1 j_2 m_1 m_2 \vert j m> <j l_f m m_f \vert J M > \nonumber \\
&& \cdot \vert \vec k \vert Y_{l_f m_f}(\hat k)
< V M \vert \vert {\cal O}_{A2} \vert \vert N \bar N (I J ) >
\label{a2def}
\end{eqnarray}
with the reduced matrix element defined as
\begin{equation}
< V M \vert \vert {\cal O}_{A2} \vert \vert N \bar N (I J ) >
= F(k) < IJ \to VM >_{SF} {\cal B}(I,J)~.
\label{reduced}
\end{equation}
The atomic $p\bar p$ state is specified by isospin component I and
total angular momentum $J=0,1$,
the latter values corresponding to the $^1S_0$ and $^3SD_1$ states
respectively.
The two-meson state $VM$ is specified by the intrinsic spin $j_{1,2}$,
the total spin coupling $j$, the relative orbital angular momentum
$l_f =1 $ and the relative momentum $\vec k$.
Eq. (\ref{reduced}) includes a final state form factor $F(k)$,
the spin-flavor weight $< IJ \to VM >_{SF}$ and an initial state interaction
coefficient ${\cal B}(I,J)$, containing the distortion in the protonium
state J with isospin component I.
Detailed expressions for these factors are summarized in Appendix \ref{appA}.

For the process $V \to \gamma $ (Fig. 2b), where the outgoing photon with
energy
$k^0$ is on-mass shell, we obtain, with the details shown in Appendix
\ref{appB}:
\begin{equation}
T_{V\to \gamma } = \vec  \epsilon \cdot \vec S (m_1) ~ Tr ( Q \varphi_V ) ~
{ e~ m_{\rho}^{3/2} \over (2k_0)^{1/2} f_{\rho} }
\end{equation}
where $\vec \epsilon $ and $\vec S (m_1) $, with projection $m_1$,
are the polarization vectors of $\gamma $
and V, respectively.
The flavor dependence of the transition is contained in the factor
$Tr (Q \varphi_V )$, where Q is the quark charge matrix and
$\varphi_V$ the $Q\bar Q$ flavor
wave function of vector meson V.
In setting up the two-step process $N\bar N (IJ) \to V M \to \gamma M$
we use time-ordered perturbation theory with the resulting matrix
element \cite{Pilkuhn}
\begin{equation}
T_{N\bar N(I J) \to V M \to \gamma M } =
\sum_{m_1}~
T_{V\to \gamma }~
{2 m_V \over m_V^2 - s} ~
T_{N\bar N(I J) \to V M}
\end{equation}
where the relativistic propagator for the intermediate vector meson
in a zero width approximation is included.
We resort to a relativistic prescription of the vector meson, since,
with the kinematical constraint $\sqrt{s} =0$, V has to be treated as a virtual
particle, which is severely off its mass-shell.
Accordingly, an additional factor $2m_V$, with the vector meson mass $m_V$,
has to be included to obtain the proper normalization.
From redefining
\begin{equation}
T_{V\to \gamma }{2 m_V \over m_V^2 -s }
\equiv
\vec \epsilon \cdot \vec S (m_1) ~A_{V\gamma}
\end{equation}
we generate the standard VDM expression of
\begin{equation}
T_{N\bar N(I J) \to V M \to \gamma M } =
\sum_{m_1}
T_{N\bar N(I J) \to V M} ~\vec \epsilon \cdot \vec S (m_1)~ A_{V \gamma }~.
\end{equation}
The VDM amplitude $A_{V \gamma }$, derived in the quark model, is:
\begin{equation}
A_{V \gamma } = \sqrt{2} ~ Tr( Q \varphi_V ) \sqrt{m_V \over k^0}
{e \over f_{\rho} } ~,
\end{equation}
which in the limit $m_v \approx k^0$ reduces to the well-known
results of \cite{Sakurai69}
\begin{equation}
A_{\rho \gamma } = e/f_{\rho} = 0.055 ~~{\rm and}~~A_{\omega \gamma}
= {1\over 3} A_{\rho \gamma } ~.
\end{equation}
The phase structure of $A_{V\gamma}$, as determined by $\varphi_V$,
is consistent with the corresponding definitions
for the strong transition matrix element.

In the radiative annihilation amplitude, the coherent sum of
amplitudes for $V= \rho$ and $\omega $, arising from different isospin
channels, has to be taken.
This gives:
\begin{equation}
T_{N\bar N (J)\to \gamma M} =
\sum_{V= \rho , \omega } \delta \cdot T_{N\bar N (IJ)\to V M \to \gamma M}
\end{equation}
where $\delta =1 $ for $V\neq M$ and $\delta =\sqrt{2}$ for $V=M$.
The additional $\delta $ accounts for the two possible contributions
to the amplitude from an intermediate state with $V=M$,
including a Bose-Einstein factor.
For the decay width of $N\bar N\to \gamma X$ we write
\begin{equation}
\Gamma_{N\bar N (J) \to \gamma X} =
2 \pi \rho_f \sum_{M, \epsilon_T , m_2 }
{1 \over (2J+1)} \vert T_{N\bar N (J)\to \gamma M} \vert ^2
\end{equation}
$\rho_f $ is the final state density and the sum is over the final state
magnetic quantum numbers of meson X ($m_2$) and of the photon (with transverse
polarization $\epsilon_T$).
The corresponding branching ratio B is:
\begin{equation}
B(\gamma X ) = { (2J+1) \over 4 \Gamma_{tot}(J) }
\Gamma_{N\bar N (J) \to \gamma X}
\end{equation}
where a statistical weight of the initial protonium state J with decay width
$\Gamma_{tot} (J)$ is taken.
With the details of the evaluation indicated in Appendix \ref{appC}, we finally
obtain for the branching ratios of $p\bar p \to \gamma X $
($X= \pi^0, \eta , \eta^{\prime}$):
\begin{eqnarray}
 B( \gamma \pi^0) &=& {3\over 4 \Gamma_{tot}(J=1)}
f( \gamma , \pi^0 ) A_{\rho \gamma}^2 \nonumber \\
&&\cdot \vert {\cal B}(0, 1) <^{13}SD_1 \to \rho^0 \pi^0 >_{SF} +
{1 \over 3}{\cal B}(1, 1) <^{33}SD_1 \to \omega \pi^0 >_{SF}
\vert^2 ~.
\end{eqnarray}
Alternatively, $B(\gamma \pi^0)$ can be expressed in terms of
the branching ratios $B(V\pi^0)$ for the strong
transitions $N\bar N \to V \pi^0$ (Eq. (\ref{A14}) of Appendix \ref{appA}):
\begin{equation}
 B( \gamma \pi^0) = { f(\gamma , \pi^0 ) \over f( V , \pi^0 )}
A_{\rho \gamma }^2 
 \left( B(\rho^0 \pi^0 ) + {1\over 9} B(\omega \pi^0 )
+ {2 \over 3} cos \beta_{J=1} \sqrt{B(\rho^0 \pi^0 )B(\omega \pi^0 )}\right)
\label{branpg}
\end{equation}
with the interference phase $\beta_{J=1}$ determined by
\begin{equation}
cos \beta_{J=1} = {Re \left\{ {\cal B}(0,1)^{\ast}{\cal B}(1,1) \right\}
\over \vert {\cal B}(0,1){\cal B}(1,1) \vert }~.
\label{interf1}
\end{equation}
The same equations apply for $X= \eta$ and $\eta^{\prime}$ with $\pi^0$ being
replaced by the respective meson.
Here, a kinematical phase space factor $f$ is introduced, which can
be identified with those derived in specific models
(Eqs. (\ref{A13}) and (\ref{C10})) or taken from phenomenology.
Values for the branching ratios on the right hand side of Eq. (\ref{branpg})
can either
be taken directly from experiment or determined in the quark model
considered in Appendix \ref{appA}.
Magnitude and sign of the interference term, as determined by
$cos \beta_{J=1}$, solely depends on initial state interaction for the
spin-triplet $N\bar N$ state (J=1), as expressed by the coefficients
${\cal B}(I, J=1)$.

Similarly, for the branching ratios of $p\bar p \to \gamma X$ $(X=\rho^0 ,
\omega , \gamma )$, now produced from the spin-singlet state (J=0) of
protonium, we obtain:
\begin{equation}
 B( \gamma \rho^0) = { f(\gamma , \rho^0 ) \over f( V , V)}
A_{\rho \gamma }^2
 \left( {1\over 9} B(\rho^0 \omega ) + {2} B( \rho^0 \rho^0 )
+ {2 \sqrt{2} \over 3} cos \beta_{J=0} \sqrt{B(\rho^0 \omega )B(\rho^0 \rho^0)
}\right) ~,
\label{branrg}
\end{equation}
\begin{equation}
B( \gamma \omega ) = { f(\gamma , \omega ) \over f( V , V)}
A_{\rho \gamma }^2
 \left( B(\rho^0 \omega ) + {2\over 9} B( \omega \omega )
+ {2 \sqrt{2} \over 3} cos \beta_{J=0} \sqrt{B(\rho^0 \omega )
B(\omega \omega ) }\right) ~,
\label{branog}
\end{equation}
and
\begin{eqnarray}
B( \gamma \gamma )  = { f(\gamma , \gamma ) \over f( V , V)}
A_{\rho \gamma }^4&
\left\{ B(\rho^0 \rho^0 ) + {2\over 9} B( \omega \rho^0) +
{1 \over 81} B( \omega \omega ) +
+{2 \over 9} \sqrt{ B( \rho^0 \rho^0 ) B( \omega \omega )} +
\right.
\nonumber \\
& \left. + {2 \sqrt{2} \over 3} cos \beta_{J=0} \sqrt{B(\rho^0 \omega )}
\left( \sqrt{B(\rho^0 \rho^0 )} + {1\over 9} \sqrt{B(\omega \omega )}
\right) \right\}
\label{brangg}
\end{eqnarray}
with the interference determined as
\begin{equation}
cos \beta_{J=0} = {Re \left\{ {\cal B}(0,0)^{\ast}{\cal B}(1,0) \right\}
\over \vert {\cal B}(0,0){\cal B}(1,0) \vert }~.
\label{interf0}
\end{equation}
Again, the sign and size of the interference $cos \beta_{J=0}$ are fixed by the
the initial state interaction, here for protonium states with J=0.

Eqs. (\ref{branpg}), (\ref{branrg}) and (\ref{branog}) are analogous to
those of Ref. \cite{Delcourt82};
this is also true for Eq. (\ref{brangg}) in the SU(3) flavor limit
with $B(\rho^0 \rho^0)= B(\omega \omega )$.
However, the essential and new feature of the present derivation is that
the interference term is completely
traced to the distortion in the initial protonium state.
The possibility to link the interference terms $cos \beta_J$ to the initial
state interaction in protonium is based on the separability of the
transition amplitude $T_{N \bar N (IJ) \to V M }$.
The sign and size of $cos \beta_J$ (J=0,1) will have a direct physical
interpretation, which will be discussed in the following chapter.

We briefly comment on alternative model descriptions for the strong transition
amplitudes $N\bar N \to VM$ and its consequences for the interference terms
in radiative $p\bar p$ decays.
Competing quark model approaches in the description of $N\bar N$ annihilation
into two mesons concern rearrangement diagrams as opposed to the planar diagram
of the A2 prescription of Fig. 2a.
In the rearrangement model a quark-antiquark pair of the initial $N\bar N$ state
is annihilated and the remaining quarks rearrange into two mesons.
The quantum numbers of the annihilated quark-antiquark pair are either that
of the vacuum ($^3P_0$-vertex, R2 model \cite{Green84}) or that of a gluon
($^3S_1$-vertex, S2 model \cite{Maruy85,Henley86}).
In the R2 model, two ground state mesons cannot be produced from an initial
$N\bar N$ state in a relative S-wave;
hence R2 is not applicable to the annihilation process considered here.
The S2 model generates transition matrix elements for $p\bar p \to VM$,
which are analogous to the ones of the A2 model of Eqs. (\ref{a2def})
and (\ref{reduced}), but with different absolute values for the spin-flavor
weights $<IJ \to VM>_{SF}$ \cite{Maruy85,Henley86}.
However, the relative signs of the matrix elements $<IJ \to \rho M>$ and
$<IJ \to \omega M>$ are identical to the ones of the A2 model, except in the
case $M=\eta$ where it is opposite.
Therefore, results for branching ratios $B(\gamma M)$ of radiative decays
expressed in terms of the branching ratios $B(VM)$ are identical both in the
A2 and the S2 approach, except for $B(\gamma \eta)$ where $cos \beta_{J=1}$
changes sign.
But, as will be shown later, the sign structure of $cos \beta_J$ deduced in
the framework of the A2 model is consistent with the one deduced from
experiment.

Possible deviations from the formalism presented here include contributions from
virtual $N\bar \Delta \pm \Delta \bar N$ and $\Delta \bar \Delta$ states
to the annihilation amplitudes as induced by initial state interaction.
The role of $\Delta$ state admixture and its effect on $p\bar p$ annihilation
cross sections in the context of quark models was studied in
Refs. \cite{Maruy87,GreenNis}.
Although contributions involving annihilation from $N\bar \Delta$ and $\Delta
\bar \Delta$ states can be sizable \cite{GreenNis}, the overall effect
on the annihilation cross section is strongly model dependent.
In the case of the A2 model \cite{Maruy87}, these contributions are found to
be strong for $N\bar N$ D-wave coupling to channels with a virtual $\Delta$ in
the S-wave, hence dominantly for the $^{13}SD_1$ partial wave, where for
isospin $I=0$ the tensor force induces strong mixing.
However, for the radiative decay processes at rest considered here, the
possible $N\bar \Delta \pm \Delta \bar N$ configurations only reside in the
$^{33}SD_1$ state (here $^{33}SD_1 \to \pi^0 \omega$ and
$^{33}SD_1 \to \eta \rho^0$).
Due to the weak D-wave coupling in the I=1 channel, $N\bar \Delta$
configurations play a minor role and are neglected.

Alternatively, the strong transition amplitudes $N\bar N \to VM$ can be
derived in baryon exchange models \cite{Moussa84,Hipp91,Mull95a}.
Here however, the analysis is strongly influenced by the presence both
of vector and tensor couplings of the vector mesons to the nucleon,
by contributions of both $N$ and $\Delta$ exchange (where the latter one
contributes to the $\rho^0 \rho^0$ and $\pi^0\rho^0$ channels) and by the
addition of vertex form factors.
The interplay between these additional model dependencies complicates an
equivalent analysis.
Due to simplicity we restrict the current approach to a certain class of quark
models, although deviations from the analysis given below when applying
baryon exchange models cannot be excluded.

\section{PRESENTATION OF RESULTS}
\label{results}

In Sec. \ref{branch} we discuss the direct application of the quark model
approach
to the radiative $N\bar N$ annihilation process.
In Sec. \ref{isospint} we focus specifically on the isospin interference effects
occuring in radiative transitions.
We show that the the interference term is solely determined
by the isospin dependent $N \bar N$ interaction,
and give theoretical predictions for the phase
$cos \beta_J$ in various $N\bar N$ interaction models.
Sign and size of $cos \beta_J$ can be interpreted by the dominance of
either the $p\bar p$ or the $n\bar n$ component of the protonium
wave function in the annihilation region.
Furthermore we show that extraction of the interference term from
experimental data is greatly affected by the choice of the kinematical
form factor.
Finally we comment on the applicability of the vector dominance approach
to the $p\bar p \to \gamma \phi$ transition.

\subsection{Branching ratios of radiative protonium annihilation}
\label{branch}

In a first step we directly evaluate the expression for $B(\pi^0\gamma)$
and $B( X \gamma)$, $X=\eta ,~\omega , ~\eta^{\prime} ,~\rho$ and $\gamma $,
as given by Eq. (\ref{branpg}), (\ref{branrg}) - (\ref{brangg})
and Eq. (\ref{A14}).
To reduce the model dependencies we choose a simplified phenomenological
approach as advocated in studies for two-meson branching ratios in $N\bar N$
annihilation \cite{Dover91}.
The initial state interaction coefficients ${\cal B}(I,J)$
are related to the probability for a protonium state with spin J and isospin I,
with the normalization condition $\vert {\cal B}(0,J) \vert^2 +
\vert {\cal B}(1,J) \vert^2 =1$.
The total decay width of state J is given by $\Gamma_{tot} (J)$ with the
separation into isospin contributions as $\Gamma_{tot} (J) = \Gamma_0 (J)
+ \Gamma_1 (J)$.
We identify the ratio of isospin probabilities $\vert {\cal B}(0,J) \vert^2
/ \vert {\cal B}(1,J) \vert^2$ with that of
partial annihilation widths $\Gamma_0 (J) / \Gamma_1 (J)$.
For our calculations we adopt the isospin probabilities deduced from
protonium states obtained with
the Kohno-Weise $N\bar N$ potential \cite{Kohno86},
where $p\bar p -n\bar n$ isospin mixing and tensor coupling
in the the $^3SD_1$ state are fully included \cite{Carbonell89}.
The resulting values for ${\cal B} (I,J)$ are
shown in Table \ref{tab1}.
The kinematical form factor $f( \gamma , X)$ is taken of the
form \cite{Vander88}
\begin{equation}
f(\gamma , X) = k \cdot  exp\left\{ -A \sqrt{ s - m_X^2} \right\}
\label{van}
\end{equation}
where k is the final state c.m. momentum with total energy $\sqrt{s}$.
The constant $A=1.2~GeV^{-1}$ is obtained from a phenomenological fit
to the momentum dependence of various multipion final states in
$p\bar p$ annihilation \cite{Vander88}.
Results for the branching ratios in this simple model ansatz
are given in Table \ref{tab2}.
For the decay modes $\eta \gamma$ and $\eta^{\prime} \gamma $ we use
a pseudoscalar mixing angle of $\Theta_p = -17.3^{\circ}$ \cite{Ams92}.
The model contains a free strength parameter, corresponding to the strong
annihilation into
two mesons in the two-step process. 
Since we compare the relative strengths of the branching ratios,
we choose to normalize the entry for $B(\gamma \pi^0)$
to the experimental number.
The A2 quark model prediction for the hierarchy of branching ratios
is consistent with experiment.
In particular, the relative strength of transitions from the spin-singlet
$(^1S_0)$ and triplet ($^3SD_1$) $N\bar N$ states is well
understood.
The results of Table \ref{tab2} give a first hint, that the VDM approach is a
reliable tool in analysing the radiative decays of protonium.
Furthermore, all considered branching ratios are consistent with
minimal kinematical and dynamical assumptions.
We stress that the good quality of the theoretical fit to the experimental
data of Table \ref{tab2} should not be overemphasized given the simple
phenomenological approach where initial state interaction is introduced
in an averaged fashion.
Although the A2 model provides a reasonable account of $N\bar N$
annihilation data, discrepancies remain in certain two-meson channels
\cite{Dover92,Dover91}.
In particular, observed two-meson annihilation branching ratios can show strong
deviations from simple statistical or flavor symmetry estimates
(dynamical selction rules), which in their
full completeness cannot be described by existing models.
Furthermore, theoretical predictions for two-meson branching ratios can be
strongly influenced by initial
state $N\bar N$ interaction (see for example Ref. \cite{Maruy88}),
as in the case of radiative decays,
but also by the possible presence of final state meson-meson scattering
\cite{muhm,Mull95b}.
Given these limitations in the understanding of two-meson annihilation
phenomena we will in turn dominantly focus on the determination of the
interference term present in radiative $p\bar p$ decays.
Here $N\bar N$ annihilation model dependencies are avoided by resorting
to the experimentally measured two-meson branching ratios. 

\subsection{Isospin interference and initial state interaction}
\label{isospint}

In a second step we focus on the determination and interpretation of
the isospin interference terms $cos \beta_J $ (J=0,1) given by
Eqs. (\ref{interf1}) and (\ref{interf0}),
which in turn are related to the $N\bar N$ initial state
interaction via the coefficients ${\cal B}(I, J)$ in Eq. (\ref{A10}).

A full treatment of protonium states must include both Coulomb and
the shorter ranged strong interaction, where the coupling of
$p\bar p$ and $n\bar n$ configurations is included.
The isospin content of the corresponding protonium wave function $\Psi $
depends on r; for large distances $\Psi $ approaches a pure $p\bar p$
configuration, i.e., $\Psi (I=0) =  \Psi (I=1)$.
As r decreases below 2 fm, $\Psi$ starts to rotate towards an isospin
eigenstate, i.e. $\Psi $ takes the isospin of the most attractive potential
in the short distance region.
The $N\bar N$ annihilation process under consideration here is most
sensitive to the behaviour of $\Psi $ for $r \leq 1 $ fm,
where the strong spin- and isospin dependence of the $N\bar N$ interaction
may cause either the $I=0$ or the $I=1$ component to dominate.
The consequences of the spin-isospin structure for energy shifts and
widths of low lying protonium states have been discussed in Refs.
\cite{Carbonell89,Kauf79,Richard82}.
The sensitivity of $p \bar p - n\bar n$ mixing in protonium states
to changes in the meson-exchange contributions to the $N \bar N$ interaction
was explored in \cite{Dover91}.

Let us first discuss the physical interpretation of the interference terms
$cos \beta_J $.
For a protonium state described by a pure $p\bar p$ wave function,
the isospin dependent initial
state interaction coefficients are related, ${\cal B}(I=0 , J) =
{\cal B}(I=1 , J)$.
Similarly, for a protonium state given by a pure $n\bar n$ wave function
in the annihilation region, that is $\Psi (I=0) = - \Psi (I=1)$,
${\cal B}(I=0 , J) =  - {\cal B}(I=1 , J)$.
Together with Eqs. (\ref{interf1}) and (\ref{interf0}), we obtain
for the interference terms
\begin{equation}
cos \beta _J
= \left\{ \begin{array}{*{2}{c}}
+1  & \, {\rm for~ pure~ p\bar p} \\
-1 & \, {\rm for~ pure~ n\bar n}
\end{array}\right. ~.
\end{equation}
Therefore, a dominant $p\bar p$ component in the protonium wave function leads
to constructive interference in radiative annihilation, with $cos \beta_J =1$.
Destructive interference reflects the dominance of the $n\bar n$ component
in the annihilation region of the protonium state.
Given this direct physical interpretation of the interference terms,
radiative annihilation serves as a indicator for the isospin dependence
of the $N\bar N$ protonium wave functions.

For quantitative predictions of the interference terms and for comparison,
we resort to protonium wave functions calculated \cite{Carbonell89} with
three different potential models
of the $N\bar N$ interaction,
that by Kohno and Weise
\cite{Kohno86} (KW) and the two versions of the Dover-Richard
\cite{Dover80,Richard82} (DR1 and DR2) potentials.
The calculation of Ref. \cite{Carbonell89}
takes fully account of the neutron-proton mass difference, tensor coupling
and isospin mixing induced by the Coulomb interaction.

Results for the interference terms $cos \beta _J$ as deduced from
the three different potential models are given in Table \ref{tab3}.
The value of the range parameter $d_{A2}$ in the initial state form factor
entering in Eq. (\ref{A10}) is adjusted to the range of the annihilation
potential of the respective models.
With the choice of $d_{A2} = 0.12 ~fm^2$ (KW and DR2) and
$d_{A2} = 0.03 ~fm^2$ (DR1) the calculated ratios of isospin probabilities
$\vert {\cal B}(0,J) \vert^2 / \vert {\cal B}(1,J) \vert^2$ are close to
those of partial annihilation widths $\Gamma_0 (J) / \Gamma_1 (J)$
calculated in Ref. \cite{Carbonell89}.
All three potential models consistently predict constructive interference
for radiative annihilation from the atomic $^1S_0$ state, indicating
a dominant $p\bar p$ component.
For radiative annihilation from the spin triplet state $^3S_1$
predictions range from nearly vanishing (DR1) to a sizable destructive
interference, where latter effect can be traced to a dominant
short ranged $n \bar n$ component in the protonium state.
In Table \ref{tab3} we also indicate predictions for the interference
term $cos \beta_1$, where the D-wave admixture in the $^3SD_1$ state
has been included.
The results are obtained for the specified values of $d_{A2}$ with an
additional choice of hadron size parameters (that is $R_N^2 /R_M^2 = 0.6$
or $<r^2 >^{1/2}_N / <r^2>^{1/2}_M =1.2$) entering in the expression of
Eq. (\ref{A10}).
The inclusion of D-wave admixture in the initial state interaction coefficients
${\cal B}(I,J=1)$ as outlined in Appendix \ref{appA} is a particular
feature of the A2 quark model.
Hence, predictions for $cos \beta_1$ with the $^3D_1$ component of
the atomic $^3SD_1$ state included are strongly model dependent and
should not be overestimated.
Generally, inclusion of the D-wave component in the form dictated by the
quark model tends to increase the values of the interference terms.

We also investigated the sensitivity of the interference term $cos\beta _J$
on the range of the initial state form factor, expressed by the coefficient
$d_{A2}$.
Although the absolute values for the initial state interaction coefficients
${\cal B}(I,J)$ sensitively depend on the specific value for $d_{A2}$,
variation of $d_{A2}$ by up to 50 \% has little influence on sign and also
on size of $cos \beta_J$.
Thus, predictions for the interference terms $cos \beta_J$ in all three
potential models considered, are fairly independent on the specific
annihilation range of the $N\bar N$ initial state.

The models used for describing the $N\bar N$ initial state interaction
in protonium are characterized by a state independent, complex
optical potential due to annihilation.
Potentials of this type reproduce the low-energy $p\bar p$ cross sections
and protonium observables, such as energy shifts and widths, fairly well.
A more advanced fit \cite{pignone}
to $N\bar N$ scattering data, in particular to the
analysing powers for elastic and charge-exchange scattering, requires
the introduction of an explicit state and energy dependence in the
phenomenological short range part of the $N\bar N$ interaction.
At present, latter $N\bar N$ potential \cite{pignone}
was not applied to the protonium
system; hence the model predictions of Table \ref{tab3} should be regarded
as a first estimate for the $p\bar p -n\bar n$ mixing mechanism in the
$N\bar N$ annihilation region.

\subsection{Isospin interference from data}
\label{isospind}

The VDM approach allows to relate the branching ratios of radiative
annihilation modes to branching ratios with final states containing one or two
vector mesons.
Using these measured branching ratios in Eqs. (\ref{branpg}) and
(\ref{branrg}) - (\ref{brangg}) we can
extract the interference terms $cos \beta_J$ directly
from experiment.
However, conclusions on the sign and size of the interference terms
strongly depend on the choice of the kinematical form factor
$f(X_1, X_2)$, $X_1$ and $X_2 = \gamma$ or meson, entering in the different
expressions.
A first analysis \cite{Ams93} for determining the interference terms from data
was performed by the Crystal Barrel Group, assuming a form factor of
\cite{Hippel72}
\begin{equation}
f(X_1 , X_2 ) = k \left( { (kR)^2 \over 1 + (kR)^2 } \right) ~,
\label{hippel}  
\end{equation}
where k is the final state c.m. momentum and the interaction range is chosen
as $R = 1.0 ~fm$.
This form factor is appropriate for small momenta k,
taking into account the centrifugal barrier effects near threshold.
However, for radiative decays, with high relative momenta in the final state,
the choice of Eq. (\ref{van}) is more appropriate,
it contains an exponential which restricts
the importance of each decay channel to the energy region near threshold.
This can be regarded as a manifestation of multichannel unitarity,
that is the contribution of a given decay channel cannot grow linearly with
k (as in the form of Eq. (\ref{hippel})),
since other channels open up and compete for
the available flux, subject to the unitarity limit.
Also, the latter form factor is given a sound phenomenological basis in
$N\bar N$ annihilation analyses, for a more detailed discussion see for
example Ref. \cite{Amsler97}.
Extracted values for the interference terms $cos \beta _J$ with
different J and different prescriptions for the kinematical form factor
are given in Table \ref{tab4}.
We also include there a third choice for the
kinematical form factor (Eq. (\ref{A13})), as deduced from the A2 quark model
description of the $N\bar N$ annihilation process.
Although finite size effects of the hadrons are included here,
through the harmonic oscillator ansatz for the hadron wave functions
the form factor is again useful for low relative momenta k.
For the results of Table \ref{tab4} we use the measured branching ratios of
$p\bar p \to \pi^0 \rho^0$ \cite{Chiba88}, $\pi^0 \omega, ~\eta \omega,
~\omega \omega , ~\eta^{\prime} \omega$ \cite{Amsler93},
$\eta \rho$ \cite{Adiels89} or \cite{Amslermyhrer91}, $\rho \omega $
\cite{Bizarri69} and $\eta^{\prime} \rho$ \cite{Amsler97}.
Values for $cos \beta_J$ using the phase space factor of Eq. (\ref{hippel})
are directly taken from the original analysis of Ref. \cite{Ams93}.
Error estimates for the other entries in Table \ref{tab4} assume statistical
independence of the measured branching ratios.
For the radiative decay channel $\eta^{\prime} \gamma $ only an upper limit
for $cos \beta_1$ can be given.

For all three choices of the kinematical form factor, the extracted values of
$cos \beta _J$ are consistent with the VDM assumption as they correspond
to physical values.
However, as evident from Table \ref{tab4}, conclusions on sign and size of the
interference strongly depend on the form of the kinematical phase space factor.
For the preferred choice, i.e. Eq. (\ref{van}), we deduce destructive interference
for radiative annihilation from the $^3SD_1$ state, while for the $^1S_0$ state
the corresponding isospin amplitudes interfere constructively.
This is in contrast to the original analysis of Ref. \cite{Ams93}, where the
interference term is determined to be almost maximally destructive
for all channels considered.
Given the large uncertainties for $cos \beta _J$ using the preferred
form factor,
the values deduced from data are at least qualitatively consistent
with the theoretical predictions of Table \ref{tab3}, indicating a
dominant $p\bar p$ component for the $^1S_0$ and a sizable $n\bar n$ component
for the $^3SD_1$ protonium wave function.
As discussed in Sec. \ref{isospint}, precise values for $cos\beta_J $ are rather
sensitive on the isospin decomposition of the protonium wave function in the
annihilation region.
However, the current uncertainties in the experimental data should be
very much improved to allow a more quantitative
analysis of the isospin dependence of the $N\bar N$
interaction.

\subsection{Vector dominance model and the $p\bar p \to \gamma \phi$
transition}
\label{phi}

Measurements on nucleon-antinucleon annihilation reactions into
channels containing $\phi$ mesons indicate apparent violations
of the Okubo-Zweig-Iizuka (OZI) rule \cite{Amsler97}.
According to the OZI rule, $\phi$ can only be produced through its non-strange
quark-antiquark component, hence $\phi $ production should vanish for an
ideally mixed vector meson nonet.
Defining the deviation from the ideal mixing angle $\theta _0 =35,3^{\circ}$
by $\alpha = \theta - \theta_0$ and asssuming the validity of the OZI rule,
one obtains the theoretically expected ratio of branching ratios
\cite{Dover92}:
\begin{equation}
R(X)=B( N \bar N \to \phi X )/B(N \bar N \to \omega X) = tan^2 \alpha
\approx 0.001 - 0.003
\label{ratiox}
\end{equation}
where X represents a non-strange meson or a photon.
Recent experiments \cite{Amsler97} have provided data on the
$\phi /\omega $ ratios
which are generally larger than the standard estimate of Eq. (\ref{ratiox}).
The most notable case is the $\phi \gamma $ channel
for $p\bar p$ annihilation in liquid hydrogen \cite{Ams95}, where data
show a dramatic violation of the OZI rule of up to two orders of
magnitude, that is $R(X = \gamma ) \approx 0.3$.
Substantial OZI rule violations in the reactions $p\bar p \to X \phi$
can possibly be linked to the presence of strange quark components
in the nucleon \cite{Ellis95,Gutsche97}.
However, apparent OZI rule violations can also be generated by conventional
second order processes, even if the first order term corresponds to a
disconnected quark graph \cite{Gortch96,Marku97}.

In Refs. \cite{Marku97,Locher94} the apparently large value for the
branching ratio $B(\gamma \phi)$
for $p\bar p$ annihilation in liquid hydrogen is explained within
the framework of the VDM.
Using the experimental rates of $B(\rho \phi) =(3.4 \pm 1.0)
\times 10^{-4}$ and $B(\omega \phi) =(5.3 \pm 2.2) \times 10^{-4}$
\cite{Reifen91} as inputs, the branching ratio $B(\gamma \phi )$
is given in the VDM by:
\begin{equation}
B(\gamma \phi ) = {f(\gamma , \phi ) \over f(\omega V)}
\left( 12.0 + cos\beta_0 \cdot 8.5 \right) \cdot 10^{-7}
\label{phirel}
\end{equation}
Since the $\phi \omega $ and $\phi \rho $ channels also violate
the OZI rule estimate, $R(X= \rho) \approx R(X=\omega ) \approx 10^{-2}$
\cite{Amsler97},
the standard $\omega -\phi$ mixing cannot be the dominant
mechanism for the production of the $\phi \omega $ and $\phi \rho $ channels
and 
the formalism developed in Sec. \ref{form} cannot be used to determine
the phase structure of the interference term $cos \beta_0$ for
$B(\gamma \phi)$.
Consequently, the interference term $cos \beta_0$ extracted in the
$\gamma \omega $ reaction is not necessarily consistent with that of
the $\gamma \phi$ decay channel.
For maximal constructive interference $(cos\beta = 1)$ one obtains
an upper limit for $B(\gamma \phi)$ in the VDM calculation of:
\begin{eqnarray}
 B( \gamma \phi) &=& 2.7 \times 10^{-5} \quad {\rm for~}
f=k^3~\cite{Locher94} \nonumber \\
 B( \gamma \phi) &=& 1.5  \times 10^{-6} \quad {\rm for~}
f~ {\rm given ~ by~ Eq.~} (\ref{van})
\label{upper}
\end{eqnarray}
This is to be compared with
the experimental result $B(\gamma \phi) = (2.0 \pm 0.4)\times 10^{-5}$
\cite{Ams95}.
The possibility to explain the experimental value of $B(\gamma \phi)$ in VDM
depends again strongly on the choice of the kinematical form factor.
In Ref. \cite{Locher94} the form $f=k^3$ is used, appropriate for relative
momenta $k$ near threshold, resulting in an upper limit which lies slightly
above the observed rate for $B(\gamma \phi)$.
With the choice of Eq. (\ref{van}) the upper value underestimates
the experimental number by an order of magnitude.
When we extract the interference terms $cos \beta_J$ from the conventional
radiative decay modes with the choice $f= k^3$, we obtain:
$cos \beta_1 = -1.32 $ for $\pi \gamma$, $cos \beta_1 =
-0.94 $ for $\eta \gamma$ and $cos \beta_0 =-.90 $ for
$\omega \gamma $.
Hence, a near threshold prescription for the kinematical form factor
in the VDM
leads to maximal destructive interference for all channels
considered,
exceeding even the physical limit in the case of $\pi \gamma $.
This would indicate a nearly pure $n\bar n$ component in the annihilation
range of the protonium wave functions for both the J=0 and 1 states.
These results are in strong conflict with the theoretical expectations
for $cos \beta_J$ reported in Sec. \ref{isospint},
where at least
qualitative consistency is achieved with the kinematical form factor
of Eq. (\ref{van}).

Recent experimental results \cite{evang} for the reaction cross section
$p\bar p \to \phi \phi$ exceed the simple OZI rule estimate by about two orders
of magnitude.
Therefore, in the context of VDM an additional sizable contribution to the
branching ratio $B(\gamma \phi)$ might arise, although off-shell,
from the $\phi\phi$ intermediate state.
With an estimated cross section of $p\bar p \to \omega \omega $ of about
5 mb in the energy range of the $\phi \phi$ production experiment, the ratio
of cross sections is given as $\sigma_{\phi \phi}/\sigma_{\omega \omega}
\approx 3.5~\mu b / 0.5 ~mb$ \cite{evang}.
Given the measured branching ratios of $\omega \omega $ \cite{Amsler93}
and $\omega \phi$ \cite{Reifen91} we can simply estimate the ratio of
strong transition matrix elements for annihilation into $\phi \phi$ and
$\omega \phi$ from protonium of $\sqrt{B(\phi \phi )/B( \omega \phi )}
\approx 0.43 $.
For this simple order of magnitude estimate we assume that $\sigma_{\phi \phi}/
\sigma_{\omega \omega}$ is partial wave independent and phase space corrections
are neglected.
With the VDM amplitude $A_{\phi \gamma} ={ \sqrt{2} \over 3} A_{\rho \gamma}$
we obtain an upper limit of $B(\gamma \phi) \approx  2.3 \times 10^{-6}$ with
f given by Eq. (\ref{van}), where the contribution of the $\phi \phi $
intermediate state is now included.
Excluding an even further dramatic enhancement of the $\phi \phi$ channel
for $N\bar N$ S-wave annihilation, inclusion of the $\phi\phi$ intermediate
state does not alter the conclusions drawn from the results of Eq.
(\ref{upper}).
Hence,
the large observed branching ratio for $\gamma \phi$ remains unexplained
in the framework of VDM.

\section{SUMMARY AND CONCLUSIONS}
\label{sum}

We have performed a detailed analysis of radiative $p\bar p$ annihilation
in the framework of a two-step process, that is $p\bar p$ annihilates
into two-meson channels containing a vector meson which is subsequently
converted into a photon via the VDM.
Both processes are consistently formulated in the quark model, which allows
to uniquely identify the source of the isospin interference present in
radiative transitions.
Based on the separability of the transition amplitude $N\bar N \to VM$,
sign and size of the interference terms can be linked to the dominance
of either the $p\bar p$ or the $n\bar n$ component of the 1s protonium wave
function in the annihilation region, hence constitutes a direct test
of the isospin dependence of the $N \bar N$ interaction.  

In a first step we directly applied the quark model in a simplified
phenomenological approach to the radiative $N\bar N $ annihilation process.
Model predictions are consistent with data and confirm the usefulness
of VDM in the analysis of radiative transitions.
In a second step we discussed sign and size of the interference term as
expressed by $cos \beta_J$ ($J=0,1$).
Direct predictions of $cos \beta_J$, as calculated for different
potential models of the $N\bar N$ interaction, are qualitatively consistent,
in that a sizable constructive interference is deduced for radiative
annihilation from the atomic $^1S_0$ state, while for the $^3S_1$ state the
interference term is vanishing or destructive.
These predictions should be tested with more realistic parameterizations
of the $N\bar N$ interaction \cite{pignone}.
Extraction of the interference effect from data is greatly influenced by
the choice of the kinematical form factor associated with the transition.
Values of $cos \beta_J$ determined for the preferred form of Eq. (\ref{van})
are qualitatively consistent with our theoretical study;
however, a more quantitative analysis is restricted by the present
uncertainties in the experimental data.
Within the consistent approach emerging from the analysis of non-strange
radiative decay modes of protonium, an explanation of the measured
branching ratio for the OZI suppressed reaction $p\bar p \to \gamma \Phi$
cannot be achieved.   
New mechanisms, linked to the strangeness content in the
nucleon, may possibly be responsible for the dramatic violation of the
OZI rule in the $\gamma \Phi$ final state.

\begin{acknowledgments}
This work was supported by a grant of
the Deutsches Bundesministerium
f\"ur Bildung und Forschung (contract No. 06 T\"u 887) and by
the PROCOPE cooperation project (No. 96043).
We also acknowledge the generous help of Jaume Carbonell for providing
us with the protonium wave functions used in this paper.
\end{acknowledgments}

\newpage
\begin{appendix}

\section{Nucleon-antinucleon annihilation into two mesons in the
quark model}
\label{appA}

In describing the annihilation process of $N\bar N \to VM$ where $V=\rho,~
\omega$ and $M=\pi^0,~\eta,~\rho,~\omega$ and $\eta^{\prime}$
we use the A2-model of Fig. 2a.
Detailed definitions and derivation of this particular quark model
are found in Refs. \cite{Dover92,Maruy87}.
The initial state $N\bar N$ quantum numbers are defined by $i={ILSJM}$
(I is the isospin, L is the orbital angular momentum,
S is the spin and J
is the total angular momentum with projection M).
For the final two meson state $VM$ we specify the angular momentum quantum
numbers, with $j_{1,2}$ indicating the spin of mesons 1 and 2, $j$ the total
spin coupling and $l_f$ the relative orbital angular momentum.
For the transitions of interest the quantum numbers are restricted to
L=0 and 2, corresponding to $p\bar p$ annihilation at rest in liquid hydrogen,
$j_1 =1$, representing the
vector meson, and $l_f=1$, given by parity conservation.
Taking plane waves for the initial and final state wave functions with
relative momenta $\vec p$ and $\vec k$, respectively, the transition matrix
element is given in a partial wave basis as:
\begin{eqnarray}
T_{N\bar N(i) \to V M} & = &< V (j_1) M(j_2) l_f \vert
 {\cal O}_{A2} \vert N\bar N (i)> \nonumber \\
&=& \sum_j < j_1 j_2 m_1 m_2 \vert j m> <j l_f m m_f \vert J M > \nonumber \\
&& \cdot \vert \vec k \vert Y_{l_f m_f}(\hat k) Y_{LS}^{JM~\dagger} (\hat p)
 <VM (j_1, j_2 ,j, l_f)\vert \vert {\cal O}_{A2} \vert
\vert N\bar N (i)>.
\end{eqnarray}
The reduced matrix element of the two-meson transition is given in the
A2 model as:
\begin{equation}
<VM\vert \vert {\cal O}_{A2} \vert \vert N\bar N (i)> =
F_{L,l_f} p^L exp(-d_{A2}(3/4 k^2 + 1/3 p^2) <i\to VM>_{SF} ~.
\label{A2}
\end{equation}
The factor $F_{L,l_f}$ is a positive geometrical constant depending on the size
parameters of the hadrons for given orbital angular momenta $L$ and $l_f$.
The exponentials arise from the overlap of harmonic oscillator wave functions
used for the hadrons with the coefficient $d_{A2}$ depending on the size
parameters $R_N$ and $R_M$ of the nucleon and meson:
\begin{equation}
d_{A2} = {R_N^2 R_M^2 \over 3R_N^2 + 2 R_M^2}~.
\end{equation}
The matrix elements $<i\to VM>_{SF}$ are the spin-flavor weights
of the different transitions listed in Table \ref{tab5}.
Note that with the flavor part of the vector mesons defined as
\begin{equation}
\rho^0 = {1\over \sqrt{2}} (u\bar u - d\bar d) ,~~
\omega = {1\over \sqrt{2}} (u\bar u + d\bar d)
\label{A4}
\end{equation}
the matrix elements $<i \to \rho M >$ and $<i \to \omega M >$ have same
sign.
For the tensor force coupled channel $^3SD_1$ the spin-flavor matrix elements
are simply related by a proportionality factor, dependent on the isospin
channel, but independent of the $VM$ combination, that is:
\begin{eqnarray}
&F_{L=2, l_f=1} <^{2I+1, 3}D_1\to VM>_{SF}
= C(I) \cdot F_{L=0, l_f=1} <^{2I+1, 3}S_1\to VM>_{SF}~, \nonumber \\
&C(I)= \left ( {I=0\mbox{, }-\frac{2\sqrt{2}}{5} \atop
I=1\mbox{, }\mbox{ }\mbox{ }\mbox{ }\mbox{ }\frac{2\sqrt{2}}{13}} \right)
\cdot \left ( -\frac{1}{3}
\frac{(R_N^2+R_M^2)R_N^2}{3/2R_N^2+R_M^2} \right ) ~.
\end{eqnarray}

In coordinate space the protonium wave function, including tensor coupling
and isospin mixing, is written as:
\begin{equation}
\Psi_{p\bar p} (J,S) = \sum_{L,I} \psi_{ILSJ} (r) Y_{LS}^{JM} (\hat r )~.
\end{equation}
Inserting this wave function into the expression
for the transition matrix element results in:
\begin{eqnarray}
T_{N\bar N(IJ) \to V M} & = &
\sum_j < j_1 j_2 m_1 m_2 \vert j m> <j l_f m m_f \vert J M > \nonumber \\
&& \cdot \vert \vec k \vert Y_{l_f m_f}(\hat k)
F (k) <i\to VM>_{SF} {\cal B} (I,J)~, \nonumber \\
F( k )& \equiv &  exp(-d_{A2}3/4 k^2)~.
\label{A8}
\end{eqnarray}
The distortion due to initial state interaction is contained in the
coefficient ${\cal B}(I,J)$, which is simply the overlap of the isospin
decomposed protonium wave function with the effective initial form factor
arising in the transition.
By taking the Fourier transform of the initial state form factor
contained in Eq. (\ref{A2}),
these coefficients for the 1s atomic states of protonium are defined as:
\begin{eqnarray}
{\cal B}(I,J=0)&=&F_{L=0, l_f =1}\left( 2d_{A2}/3 \right)^{-3/2} \int_0^{\infty}
dr r^2 exp(-3 r^2/(4d_{A2})) \psi_{I 0 0 0} (r)~~{\rm for}~^1S_0~,
\nonumber \\ 
{\cal B}(I,J=1)&=&F_{L=0, l_f =1}\left\{ \left( 2d_{A2}/3 \right)^{-3/2}
\int_0^{\infty}
dr r^2 exp(-3 r^2/(4d_{A2})) \psi_{I 0 1 1} (r) - \right.
\nonumber \\
&& \left.  - C(I)
\left( 2d_{A2}/3 \right)^{-7/2} \int_0^{\infty}
dr r^4 exp(-3 r^2/(4d_{A2})) \psi_{I 2 1 1} (r) \right\}~~{\rm for}~
^3SD_1~.
\label{A10}
\end{eqnarray}

The partial decay width for the annihilation of a protonium state with
total angular momentum J into two mesons $VM$ is given by
\begin{equation}
\Gamma_{p\bar p\to VM} (I,J) = 2\pi { E_V E_M \over E} k
\int d\hat k \sum_{m_1 m_2 m_f}   \vert T_{N\bar N(IJ) \to V M} \vert^2
\end{equation}
where E is the total energy and $E_{V,M} =\sqrt{m_{V,M}^2 + \vec k^2}$ the
energy of the respective outgoing meson with $\vert \vec k \vert$
fixed by energy conservation.
With the explicit form of the transition amplitude of Eq. (\ref{A8}),
the partial decay width is written as:
\begin{equation}
\Gamma_{p\bar p\to VM} (I,J) = f(V,M) <i\to VM>_{SF}^2 \vert {\cal B} (I,J)
\vert^2
\end{equation}
with the kinematical phase space factor defined by:
\begin{equation}
f(V,M) = 2\pi {E_V E_M\over E} k^3 exp\left( - 3/2 d_{A2} k^2 \right) ~.
\label{A13}
\end{equation}
Taking an admixture of initial states given by their statistical weight,
the branching ratio of S-wave $p\bar p$ annihilation into the two meson
final state $VM$ is given by:
\begin{equation}
B( V M) =
B(p\bar p \to V M) = \sum_{J=0,1} {(2J+1) \Gamma_{p\bar p \to VM}
(I,J) \over 4 \Gamma_{tot} (J)} ~.
\label{A14}
\end{equation}

\newpage

\section{Vector meson - photon conversion in the quark mo\-del}
\label{appB}

The transition $V \to \gamma$ (Fig. 2b), where $V= \rho$ or $\omega $,
can be formulated
in the quark model, and related to the physical process
of $V \to e^+ e^-$.
An explicit derivation of the latter process can be found in Ref.
\cite{Yaouanc88}.
We just quote the main results necessary for the discussion of the
radiative decays of protonium.

The $Q\bar Q \gamma$ interaction is defined by the Hamiltonian
\begin{equation}
H_I = e \int d^3x j^{\mu}_{em} (\vec x ) A_{\mu} ( \vec x )
\end{equation}
with the quark current 
\begin{equation}
j^{\mu}_{em} (\vec x ) = \bar q( \vec x) Q \gamma^{\mu} q(\vec x)
\end{equation}
where $q(\vec x) $ is the quark field and $A_{\mu}(\vec x)$ the
electromagnetic field given in a free field expansion.
For emission of a photon with momentum $\vec k$, energy $k^0$
and polarization
$\epsilon_{\mu}$ from a vector meson with momentum $\vec p_V$ we obtain:
\begin{equation}
<\gamma ( \vec k , \epsilon_{\mu}) \vert H_I \vert V (\vec p_V )>
= \delta (\vec k -\vec p_V ) T_{V \to \gamma } 
\end{equation}
with
\begin{equation}
T_{V \to \gamma } = {e (2\pi)^{3/2} \over (2k^0)^{1/2} }~
\epsilon^{\ast}_{\mu}
< 0\vert j_{em}^{\mu} (\vec x = \vec 0 ) \vert V >~.
\end{equation}
For the conversion of a vector meson V into a real photon only the spatial
part of the current matrix element contributes.
Using standard techniques for the evaluation of the current matrix element
we obtain
\begin{equation}
T_{V \to \gamma } = {e \sqrt{6} \over (2k^0)^{1/2} } ~ {\vec \epsilon \cdot
\vec S}~
Tr(Q\varphi_V )~ \psi (\vec r = 0 )
\end{equation}
with the quark charge matrix Q and the polarization $\vec S$ of the vector
meson.
The $Q\bar Q$ flavor wave function $\varphi_V$ is consistently defined
as in Eq. (\ref{A4}) of Appendix \ref{appA} and contributes to the transition amplitude:
\begin{equation}
Tr(Q\varphi_V )
= \left\{ \begin{array}{*{2}{c}}
{1\over  \sqrt{2}} & \, {\rm for }~\rho^0 \\
{1\over  3 \sqrt{2}} & \, {\rm for }~\omega
\end{array}\right. ~.
\end{equation}
The spatial part of the $Q\bar Q$ wave function
at the origin $\psi (\vec r = 0 )$
is given within the harmonic oscillator description as
$\vert \psi (0) \vert^2 = ( \pi R_M^2 )^{-3/2}$,
where the oscillator parameter $R_M$ is related to
to the rms-radius as $< r^2 >^{1/2} =\sqrt{3/8} R_M$.

Extending the outlined formalism to the physical decay process $V \to
e^+ e^-$ the decay width is given as \cite{Yaouanc88}
\begin{equation}
\Gamma_{V\to e^+ e^-} = {16 \pi \alpha^2 \over m_V^2} \left\{
Tr ( Q \varphi_V )\right\}^2 \vert \psi ( 0 ) \vert^2 
\end{equation}
with $\alpha = e^2/(4\pi)$ and mass $m_V$ of the vector meson.
Latter result can be compared to the one obtained in the vector dominance
approach resulting for example in \cite{Sakurai69}
\begin{equation}
\Gamma_{\rho^0 \to e^+ e^-} = {4 \pi \over  3} {\alpha^2 m_{\rho}\over
f_{\rho}^2 }
\end{equation}
with the decay constant $f_{\rho }$.
Hence we can identify
\begin{equation}
\vert \psi ( 0 ) \vert^2 = { m_{\rho}^3 \over 6 f_{\rho}^2 }
\end{equation}
which with the experimental result of $\Gamma_{\rho^0 \to e^+ e^- } = 6.77$
yields $f_{\rho} =5.04$ or equivalently $R_M = 3.9~GeV^{-1}$, very close to the preferred value obtained
in the analysis of strong decays of mesons.
Hence, the matrix element for the conversion of a vector meson into a real
photon is alternatively written as:
\begin{equation}
T_{V \to \gamma } = {\vec \epsilon \cdot \vec S}~
Tr(Q\varphi_V ) ~ {e~ m_{\rho}^{3/2}  \over (2k^0)^{1/2} f_{\rho} } ~.
\end{equation}
\newpage

\section{Matrix elements and decay width in radiative annihilation}
\label{appC}

In the following we present details for the evaluation of the matrix element
of Eq. (6),
which is explicitly written as:
\begin{eqnarray}
T_{N\bar N(I J) \to V M \to \gamma M} & 
= & \sum_{m_1} < j_1 j_2 m_1 m_2 \vert j m> <j l_f m m_f \vert J M >
\vert \vec k \vert Y_{l_f m_f}(\hat k)
\nonumber \\
&& \cdot 
< V M \vert \vert {\cal O}_{A2} \vert \vert N \bar N (I J ) > 
\vec \epsilon \cdot \vec S (m_1) A_{V\gamma}
\label{C1}
\end{eqnarray}
where $l_f =1$ and $j=1$, for the processes considered.
The relative final state momentum $\vec k$ and the photon
polarization $\vec \epsilon $ are written in a spherical basis as:
\begin{equation}
\vert \vec k \vert Y_{1 m_f}(\hat k) =\sqrt{3 \over 4\pi}k_{m_f}
~~{\rm and}~~
\vec \epsilon \cdot \vec S (m_1) = \epsilon_{m_1}
\end{equation}
which together with Eq. (\ref{C1}) leads to the result:
\begin{eqnarray}
&T_{N\bar N(I J) \to V M \to \gamma M} = 
\sqrt{3\over 4\pi } A_{V\gamma} 
< V M \vert \vert {\cal O}_{A2} \vert \vert N \bar N (I J ) >
{i\over \sqrt{2} } 
\nonumber \\
&\cdot \left\{ \begin{array}{*{2}{c}}
 (\epsilon \times \vec k )_M & \, {\rm for}~j_2=0,~J=1~(M=\pi^0 ,~\eta ) \\
 { (-)^{m_2} \over \sqrt{3}} (\epsilon \times \vec k)_{-m_2}
& \, {\rm for}~j_2=1,~J=0~(M=\rho^0 ,~\omega ) \end{array}\right.
\end{eqnarray}
Consequently, for the process $N\bar N \to V_1 V_2 
\to \gamma_1  \gamma_2 $
the transition matrix element is determined as:
\begin{eqnarray}
T( N \bar N (I J) \to V_1 V_2 \to \gamma_1 \gamma_2 )
=\sum_{m_2} \epsilon_{m_2} (2) A_{V_2\gamma } T(N\bar N\to
V_1 V_2 \to \gamma_1 V_2) \nonumber \\
= {1\over \sqrt{4 \pi}} A_{V_1 \gamma }A_{V_1 \gamma }
< V_1 V_2 \vert \vert {\cal O}_{A2} \vert \vert N \bar N (I J ) >
{i\over \sqrt{2}} \left( \vec \epsilon (1) \times \vec k \right)
\cdot \vec \epsilon (2)
\end{eqnarray}
where $\vec \epsilon (i)$ refer to the polarization of the photon i.

The derivation of the decay widths for the radiative transitions is
examplified here for the process $N \bar N \to \gamma \pi^0 $.
The corresponding matrix element is obtained by a coherent sum of intermediate
vector meson states $\rho  $ and $\omega $ as:
\begin{eqnarray}
& T_{N\bar N(J) \to \gamma \pi^0} =
T_{^{13}SD_1 \to \rho^0 \pi^0 \to \gamma \pi^0} +
T_{^{33}SD_1 \to \omega \pi^0 \to \gamma \pi^0}
\nonumber \\
&=\sqrt{3\over 4\pi } {i\over \sqrt{2}} (\vec \epsilon \times 
\vec k)_M
\left\{ A_{\rho^0 \gamma } < \rho^0 \pi^0 \vert \vert 
{\cal O}_{A2} \vert \vert ^{13}SD_1 > +
 A_{\omega \gamma } < \omega  \pi^0 \vert \vert 
{\cal O}_{A2} \vert \vert ^{33}SD_1 > \right\}~.
\end{eqnarray}
The decay width for $N\bar N \to \gamma \pi^0 $ is then:
\begin{equation}
\Gamma_{N\bar N \to \gamma \pi^0} =
2 \pi \rho_f \sum_{\epsilon_T , M} {1\over 2J+1} \vert
T(N\bar N(J) \to \gamma \pi^0) \vert^2
\end{equation}
with the final state density
\begin{equation}
\rho_f = {E_{\pi^0} k^2 \over E_{N\bar N} } \int d\hat k ~,
\end{equation}
$\vert \vec k \vert =k $, and the sum is over the two transverse photon
polarizations
$\epsilon_T$ and the total projection M of the $N\bar N$ protonium with total
angular momentum J.
Using
\begin{equation}
\sum_{\epsilon_T , M} \int d\hat k \vert (\vec \epsilon \times \vec k )
\vert ^2 = 8 \pi k^2
\end{equation}
together with the expression for the reduced matrix
element in Eq. (2), we finally obtain:
\begin{eqnarray}
&&\Gamma_{N\bar N \to \gamma \pi^0} = \nonumber \\
&& = f(\gamma , \pi^0 ) A_{\rho \gamma }^2
\vert < ^{13}SD_1 \to \rho^0 \pi^0 >_{SF} {\cal B} (0,1) +
{1\over 3} < ^{33}SD_1 \to \omega \pi^0 >_{SF} {\cal B} (1,1) \vert^2
\end{eqnarray}
with the kinematical phase space factor defined in analogy to
Eq. (\ref{A13}) as:
\begin{equation}
f(\gamma , M) = 2\pi {E_M k^4 \over E}
exp\left( - 3/2 d_{A2} k^2 \right)~.
\label{C10}
\end{equation}

\end{appendix}

\newpage

\begin{figure}
\caption{Two-step process $N\bar N \to M V \to M\gamma $ with $V= \rho^0,
~\omega$ and
$M=\pi^0,~\eta,~\rho^0,~\omega ,~ \eta^{\prime} ,~\gamma $ for
radiative protonium annihilation.}
\label{fig1}
\end{figure}
\begin{figure}
\caption{Quark line diagrams corresponding to $N\bar N$ annihilation
into two mesons (a) and vector meson - photon conversion (b).}
\label{fig2}
\end{figure}

\newpage

\begin{table}
\caption{Isospin probabilities $\vert {\cal B} (I,J)\vert^2$.
Values are deduced from calculation of partial annihilation widths of
1s protonium with Kohno-Weise potential \protect\cite{Carbonell89}.}
\begin{tabular}{cccc}
State & $ \vert {\cal B}(0,J)\vert^2 $ & $ \vert {\cal B}(1,J)\vert^2 $&
$\Gamma_{tot} (J) ~[keV]$ \\ 
\hline
$^1S_0$ (J=0) & 0.60 & 0.40 & 1.26 \\
$^3SD_1$ (J=1) & 0.53 & 0.47 & 0.98 \\
\end{tabular}
\label{tab1}
\end{table}

\begin{table}
\caption{Results for branching ratios B for $p\bar p \to \gamma X$ with $X=
\pi^0 , ~\eta ,~\rho^0 ,~\omega , ~\eta^{\prime}$ and $\gamma $ in the
simple model estimate.
The entry for $B(\pi^0 \gamma )$ is normalized to the experimental value.
Data are taken from Ref. \protect\cite{Ams93}.}
\begin{tabular}{lcc}
Channel & $B \times 10^6$ (model) & $B \times 10^6$ (exp.) \\
\hline
$^3SD_1 \to \pi^0 \gamma $ & 44 & $44\pm 4$ \\
$^3SD_1 \to \eta \gamma $ & 14 & $9.3 \pm 1.4$ \\
$^1S_0 \to \omega \gamma $ & 68 & $68\pm 19$ \\
$^3SD_1 \to \eta^{\prime} \gamma $ & 8.3 & $\leq 12 $\\
$^1S_0 \to \gamma \gamma $ & 0.14 & $\leq 0.63$ \\
$^1S_0 \to \rho \gamma $ & 50& $---$ \\
\end{tabular}
\label{tab2}
\end{table}

\begin{table}
\caption{Isospin interference terms $cos \beta_J$
as calculated with the 1s protonium wave functions of the KW, DR1 and DR2
potential models.
Values in brackets denote the results where the $^3D_1$ component of the
atomic $^3SD_1$ state is included with admixture fixed by A2 model.}
\begin{tabular}{cccc}
 & KW  & DR1 & DR2 \\
\hline
$cos \beta_0$ & +1.00 & +0.83 & +0.63 \\
$cos \beta_1$ & -0.90 (-0.76) & +0.10 (+0.36)& -0.41 (+ 0.53)  \\
\end{tabular}
\label{tab3}
\end{table}

\begin{table}
\caption{Isospin interference terms $cos \beta_J$
as deduced from data.
The label HQ refers to the kinematical form factor of von Hippel and Quigg
as defined in Eq. (\protect\ref{hippel}).
Similarly, labels VAN and A2 refer to the form factor prescription
of Vandermeulen and of the A2 quark model, as defined in Eqs.
(\protect\ref{van}) and (\protect\ref{A13}), respectively.
The analysis for $cos\beta_J$ (HQ) is directly taken from Ref.
\protect\cite{Ams93}.
The first line of the analysis for $\eta \gamma $ is done for
$B(\eta \rho^0 ) =( 0.53 \pm 0.14) \times 10^{-2}$
\protect\cite{Adiels89},
the second line for $B(\eta \rho^0 ) = ( 0.33 \pm 0.09) \times 10^{-2}$
\protect\cite{Amslermyhrer91}.}
\begin{tabular}{lccc}
Channel  & $cos \beta_J$ (HQ) & $cos \beta_J$ (VAN) & 
 $cos \beta_J$ (A2) \\
\hline
$^3SD_1 \to \pi^0 \gamma~(J=1)$ & $-0.75 \pm 0.11$ & $-0.10 \pm 0.28$ &
$+1.00 \pm 0.38$ \\
$^3SD_1 \to \eta \gamma ~(J=1)$ & $-0.78 \pm 0.25$ & $ -0.47 \pm 0.72$ &
$-0.43 \pm 0.76$ \\
 & $-0.58 \pm 0.48$ & $ -0.17 \pm 0.90$ & $-0.12 \pm 0.96$ \\
$^1S_0\to\omega \gamma ~(J=0)$ & $-0.60 \pm 0.18$ & $+0.15 \pm 0.38$
& $-0.21 \pm 0.28 $ \\  
$^3SD_1 \to \eta^{\prime} \gamma ~(J=1)$ & $\leq -0.26 $ & $\leq  1.65$
&$\leq -0.12$\\
\end{tabular}
\label{tab4}
\end{table}

\begin{table}
\caption{Spin-flavor matrix elements $< i \to VM>_{SF}$ for the decay
$N\bar N (L=0) \to V M$ in the A2 quark model.
These are relative matrix elements obtained from Ref.
\protect\cite{Dover92}.
Here, $\eta_{ud}$ refers to the nonstrange flavor combination
$\eta_{ud}=(u\bar u + d\bar d )/\protect\sqrt{2} $.} 
\begin{tabular}{lc}
Decay channel & $< i \to VM>_{SF}$ \\
\hline
$^{11}S_0 \to \omega \omega$ & $-\sqrt{243}$ \\ 
$^{11}S_0 \to \rho^0 \rho^0$ & $-\sqrt{243}$ \\ 
$^{31}S_0 \to \omega \rho^0$ & $-\sqrt{1350}$ \\ 
$^{13}S_1 \to \pi^0 \rho^0$ & $+\sqrt{450}$ \\ 
$^{13}S_1 \to \eta_{ud} \omega $ & $+\sqrt{450}$ \\ 
$^{33}S_1 \to \pi^0 \omega $ & $+\sqrt{338}$ \\ 
$^{33}S_1 \to \eta_{ud} \rho^0$ & $+\sqrt{338}$ \\ 
\end{tabular}
\label{tab5}
\end{table}

\newpage
\centerline{Fig. 1}
\vskip 5cm
\centerline{\epsfbox{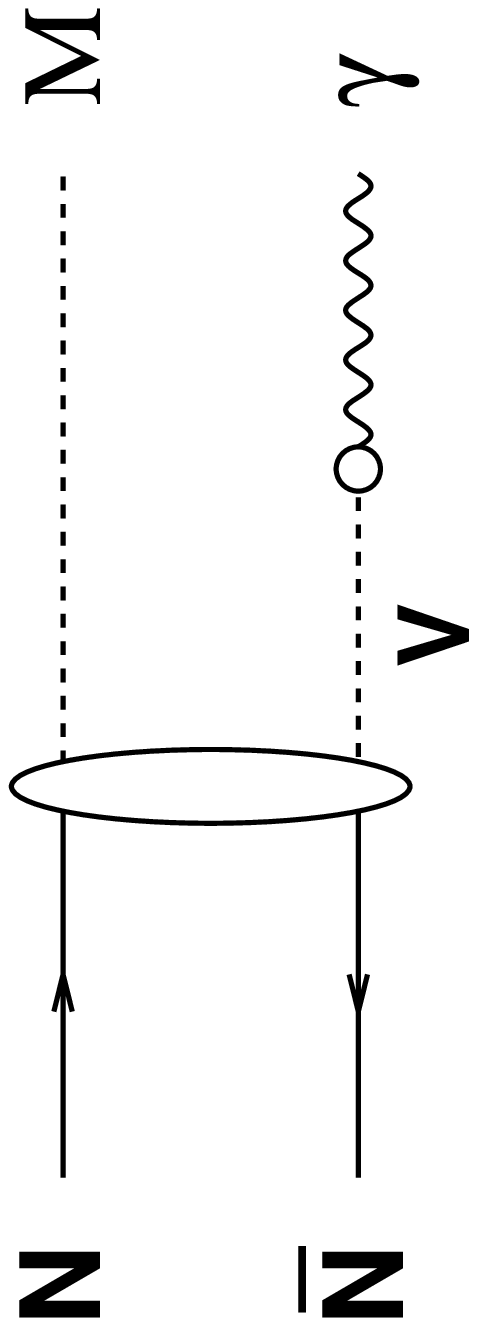}}

\newpage
\centerline{Fig. 2}
\vskip 1cm
\centerline{\epsfbox{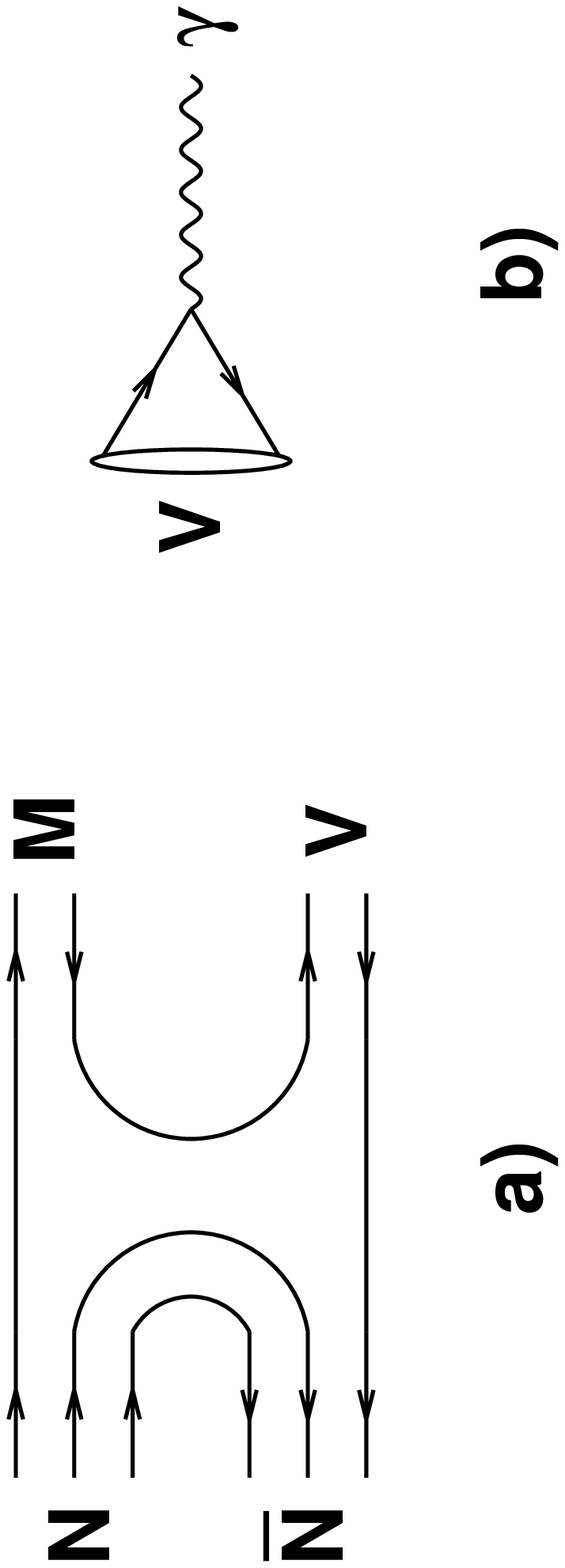}}

\end{document}